\documentstyle[twocolumn,prl,aps,epsfig]{revtex}
\newcommand{\bfr}{{\bf r}}
\newcommand{\bfk}{{\bf k}}
\newcommand{\bfK}{{\bf K}}
\newcommand{\bfq}{{\bf q}}
\newcommand{\bfx}{{\bf x}}
\newcommand{\bfy}{{\bf y}}
\newcommand{\bfz}{{\bf z}}
\newcommand{\hphi}{\hat{\phi}}

\newcommand{\ddt}{\frac{d}{dt}}

\def\jpb#1#2#3{J.~Phys.~B:~{\bf #1},\ #2\ (#3)}

\def\pra#1#2#3{Phys.~Rev.~A~{\bf #1},\ #2\ (#3)}
\def\prl#1#2#3{Phys.~Rev.~Lett.~{\bf #1},\ #2\ (#3)}
\def\sci#1#2#3{Science~{\bf #1},\ #2\ (#3)}
\def\nat#1#2#3{Nature~{\bf #1},\ #2\ (#3)}
\begin{document}
\flushbottom \draft
\title{Directional `superradiant' collisions: bosonic
amplification of atom pairs emitted from an elongated Bose-Einstein condensate}
\author{ A. Vardi and M. G. Moore}
\address{ITAMP, Harvard-Smithsonian Center for Astrophysics, Cambridge, MA 02138 \\ (\today)
\\ \medskip}\author{\small\parbox{14.2cm}{\small\hspace*{3mm}
We study spontaneous directionality in the bosonic amplification of atom pairs emitted from
an elongated Bose-Einstein condensate (BEC), an effect analogous to `superradiant' emission
of atom-photon pairs. Using a simplified model, we make analytic predictions regarding
directional effects for both atom-atom and atom-photon emission. These are confirmed by
numerical mean-field simulations, demonstrating the the feasibility of nearly perfect
directional emission along the condensate axis. The dependence of the emission angle on the
pump strength for atom-atom pairs is significantly different than for atom-photon pairs.
\\
\\[3pt]PACS numbers: 03.75.Fi, 03.75.Be, 03.75.-b }}
\maketitle

\maketitle \narrowtext

One of the most intriguing nonlinear/quantum phenomenon observed in an atomic BEC to date
has been the observation of spontaneous directionality in Rayleigh scattering of laser
light from an elongated condensate \cite{inouyet}. In this experiment a pair of condensate
momentum sidemodes were generated along with a pair of superradiant light pulses in a
cigar-shaped condensate, whereas an isotropic burst of non-condensate atoms and light was
observed from a spherical BEC.

From a quantum-optics point of view, the laser-driven BEC can be viewed as a coherent
source of correlated atom-photon pairs \cite{moore}. Various schemes have been proposed
recently in which a BEC emits correlated atom-atom rather than atom-photon pairs. One such
scheme \cite{pu,duan} makes use of microwave stark shifts to manipulate the kinematics of
spin-mixing collisions in a spinor BEC. Other schemes involve the dissociation of a BEC of
diatomic molecules . Provided a pure diatomic molecular condensate is eventually realized
\cite{timmermans,vanabeelen,javanainen,heinzen,wynar,weinstein}, its dissociation into atom
pairs could be used to generate various macroscopic entangled quantum states
\cite{YiWang01,opatrny,drummond}. A third, currently feasible, scheme proposes generating
correlated atom pairs by stimulated Raman coupling via a molecular state\cite{YouHel}.

From both fundamental and applied considerations it is important to know whether or not
spontaneous directionality is achievable in the case of atom-atom pair emission from an
elongated BEC. If a single pair of new condensates could be generated in this manner, than
by virtue of the atoms having been emitted as correlated pairs there would be strong
non-classical correlations, i.e. `squeezing', between the new condensates, with important
potential applications such as sub-shot-noise atom interferometry \cite{orzel}.

While directional effects were postulated in \cite{pu}, the analysis there relies on a
Markovian approximation in which a portion of the atomic field {\it inside} the BEC is
treated as a reservoir with an infinitely short `memory'. This approach, while justified
for a light field inside the BEC \cite{moore}, can not be applied to slowly-evolving atomic
fields, whose long `memory' is responsible for directionality. Due to this flaw, no
conclusions regarding the physics of directionality can yet be drawn. We note, however,
that there is no inconsistency with treating the atomic field {\it outside} the BEC as a
Markovian reservoir, because once atoms exit the BEC they are irreversibly lost and can no
longer influence the ensuing physics inside the BEC volume.

The purpose of this Letter is to analyze the physics of spontaneous directionality when
correlated pairs of atoms are generated from an elongated condensate; whether via
spin-mixing collisions, photodissociation of molecules, or molecular-resonance Raman
spectroscopy. We show that by introducing a large mass ratio of the paired atoms we can
reproduce the physics of superradiant Rayleigh scattering. In the limit of equal masses, on
the other hand, we smoothly map onto a new regime in which directionality still exists, but
with significant differences. While generating number-squeezed states is an important
application of `superradiant collisions', we will not analyze in detail the quantum
statistics at the present time.

We consider, for concreteness, a photodissociation scheme using two lasers to set up a
bound-free Raman transition via a far-detuned electronically excited molecular state. The
two-photon resonance is tuned into the dissociation continuum, so that excess energy goes
into the relative motion of the atom pair. Using copropagating lasers, it is possible to
transfer sufficient energy to the relative motion with negligible transfer of
center-of-mass momentum. As a variation on this scheme we note that the initial state can
also be an unbound atomic BEC state, as in \cite{YouHel}. The coupling via the molecular
excited state will transfer energy to the relative motion of atoms, again achieving the
emission of atom-atom pairs from the BEC. As experiments to implement \cite{YouHel} are
underway using a spherical BEC with counter-propagating lasers at zero detuning, only minor
changes in the laser parameters and BEC aspect ratio would be required to observe the
effects we describe. We note that our model will also describe collision-based schemes
\cite{pu,duan,YouHel} with only inconsequential adjustments.

We begin our analysis with the Hamiltonian for atom-pair generation:
\begin{eqnarray}
    \hat{H}&=&\sum_{i}\hat{H}_i +\sum_{ij}\big[\hbar \sqrt{N}\Omega_{ij} e^{i(\bfK\cdot\bfr-\Delta t)}
    \Phi(\bfr)\hphi^\dag_i(\bfr)\hphi^\dag_j(\bfr)\nonumber\\
    & &+H.c.\big].
\label{H}
\end{eqnarray}
Here, the operator $\hat{H}_i$ is the free Hamiltonian of the quantum field
$\hphi_i(\bfr)$, where $i$ describes different atomic states or species, $N$ is the
particle number, $\Omega_{ij}$ is the two-photon Rabi frequency, $\bfK$ is the two-photon
recoil momentum, and $\Delta$, {\it which supplies kinetic energy to the relative motion}
of the atom pairs, is the two-photon frequency difference minus the frequency corresponding
to the chemical potential of the molecular state and the molecular binding energy.  The
function $\Phi(\bfr)$ is the molecular BEC wavefunction, normalized to unity. For
spin-exchange collisions or molecular resonance spectroscopy $\Phi(\bfr)$ would be the
square of the atomic condensate wavefunction.

The Heisenberg equations of motion for the atomic field operators in the rotating frame
($\hphi_i(\bfr)\to\hphi_i(\bfr)\exp[-i(\delta/2)t]$)  read
\begin{eqnarray}
    \ddt\hphi_i(\bfr)&=&i\frac{\tau}{\hbar}\left[\hat{H}_i,\hphi_i(\bfr)\right]
    +i\frac{\delta}{2}\hphi_i(\bfr)\nonumber\\
    &-&i\sum_j\tau\sqrt{N}\Omega_{ij}\Phi(\bfr) e^{i\bfk\cdot\bfr}\hphi^\dag_j(\bfr).
\label{pdqs}
\end{eqnarray}
where length and time are rescaled as $\bfr\to L\bfr$ and $t\to\tau t$, with
$\tau=m_1L^2/\hbar$, $\bfk=L\bfK$ and $\delta=\tau\Delta$.

The free Hamiltonian $\hat{H}_i$ contains kinetic energy, trap potential and condensate
mean-field contributions. Atom pairs generated within the BEC volume could therefore be
described initially by a superposition of Eigenmodes of $\hat{H}_i$. The finite energy
width of these superpositions is then consistent with the finite time scale on which the
wavepacket freely evolves out of the BEC volume. Consequently, the scattering states can be
represented by a restricted basis-set of eigenstates of a `virtual cavity' within the
volume of the BEC \cite{moore}. The losses due to translational wavepacket motion can then
be modeled by a linear loss term.

Applying this approach, we use eigenstates of a box of length $L$ and width
$L/\alpha$ (with periodic boundary conditions as the lack of a physical box precludes
standing wave states) to expand $\hphi_i(\bfr)=\sum_{\bfq}\alpha
e^{i\bfq\cdot\bfr}\hat{c}_{i\bfq}$, where the states are normalized in
dimensionless units and the summation is over a grid of $\bfq$ values satisfying
$\bfq=\alpha(\ell\hat{x}+m\hat{y})+n\hat{z}$, for any integers $\ell$, $m$, and $n$. To
treat the free atomic Hamiltonians, we then approximate
\begin{equation}
   \frac{\tau}{\hbar}\left[\hat{H}_i,e^{i\bfq\cdot\bfr}\hat{c}_{i\bfq}\right]
   =\frac{1}{2}e^{i\bfq\cdot\bfr}\left[-\omega_{i\bfq}+i\gamma_{i\bfq}\right]\hat{c}_{i\bfq}
   +\hat{f}_{i\bfq}
\label{kinetic}
\end{equation}
where $\omega_{i\bfq}=\mu_i|\bfq|^2$ is the kinetic energy term  ($\mu_i=m_1/m_i$). The
decay of population from mode $\bfq$, due to translational motion of the atomic wavepacket,
is generally non-exponential. In order to approximate these losses as exponential decay we
introduce a loss rate proportional to the atomic velocity $\bfq$ over the BEC dimension
along $\bfq$ according to $\gamma_{i\bfq}=\beta\mu_i\left[(\alpha\bfq\cdot\hat{\bfx})^2
    +(\alpha\bfq\cdot\hat{\bfy})^2+(\bfq\cdot\hat{\bfz})^2\right]^{1/2}$,
where the fitting parameter $\beta=O[1]$ then depends on the exact shape of the BEC
wavefunction. Lastly, $\hat{f}_{i\bfq}$ is a standard quantum-Langevin noise operator which
must be included in any damped quantum system.

Approximating the normalized BEC ground state by the lowest energy box-eigenstate, we
arrive at a finite set of coupled Heisenberg equations
\begin{eqnarray}
    \ddt\hat{c}_{i\bfq}&=&\frac{i}{2}\left[-\mu_iq^2+\delta+i\gamma_{i\bfq}\right]\hat{c}_{i\bfq}
    -i\sum_j\chi_{ij}\hat{c}^\dag_{j(\bfk-\bfq)},
    \label{mode1}
\end{eqnarray}
where $\chi_{ij}=\tau\Omega_{ij}\sqrt{n}$, $n$ being the condensate density. We note that
we have dropped the noise operator as we are interested in the presence of instabilities
rather than in their detailed quantum statistics. Due to orthogonality, the modes
$\{\hat{c}_{i\bfq}\}$ are only coupled to the modes $\{\hat{c}_{i(\bfk-\bfq)}\}$. Equation
(\ref{mode1}) is thus a finite set of linearly coupled operator equations, whose
eigenfrequencies characterize the dynamics. The existence of an eigenfrequency with
positive imaginary part would indicate the onset of parametric instability, in which
initial spontaneously emitted atom pairs are amplified, leading to an exponential buildup
of a macroscopic mode population.

In order to demonstrate the connection between atom-pair generation and  superradiant
Rayleigh scattering, we consider first the case of a two-species atom field where $\mu_2\gg
1$ and $\chi_{ij}=\chi\delta_{i1}\delta_{j2}$, corresponding to the dissociation of a
heteronuclear molecule into one heavy and one light atom. Under these conditions, the more
massive atomic field can build up a macroscopic population while the less-massive field,
owing to its faster escape velocity, will remain only weakly populated. This situation is
analogous to the regime of BEC superradiance, with the light atoms corresponding to
scattered photons. The rapid damping of the light-atom field allows for its adiabatic
elimination, which leads to a single linear equation for the operator $\hat{c}_{1\bfq}$.
Solving this equation, we find that the mode occupation
$\langle\hat{c}^\dag_{1\bfq}\hat{c}_{1\bfq}\rangle$ grows exponentially at the rate
\begin{equation}
   G_{1\bfq}=
 \frac{4|\chi|^2\gamma_{2(\bfk-\bfq)}}
    {\left[\delta-\omega_{2(\bfk-\bfq)}\right]^2+\gamma^2_{2(\bfk-\bfq)}}
    -\gamma_{1\bfq}.
\label{Gqadelim}
\end{equation}
This expression reproduces the gain expression for superradiant Rayleigh scattering
\cite{moore} if we replace the atomic dispersion and decay terms with those of photons. The
heavy atoms, therefore, will be ejected at an angle $\hat{q}$ to the long axis satisfying
$\bfq=\bfk\pm\sqrt{\delta/\mu_2}\hat{z}$, as was observed in \cite{inouyet}. Equation
(\ref{Gqadelim}) shows that there is no saturation of directionality with increasing pump
strength $\chi$. While this is essentially the gain expression given in \cite{pu}, the
validity condition, $\mu_2\gg 1$, is not satisfied in that system.

It is interesting  to note that while it is the build up of macroscopic populations in the
slow-moving atomic modes which leads to exponential growth and directionality, it is the
losses of the fast-moving atoms or photons which is minimized. This state of affairs can be
understood if we view the light-atom (or photon) `virtual cavity' modes as intermediate
states through which the particles pass before decaying irreversibly into a reservoir of
modes. It is well known in quantum optics that the rate of coupling to a continuum is
reduced if a short-lived intermediate state is introduced, as its energy width acts as a
detuning making the transition increasingly off-resonant as the lifetime decreases.

In the remainder of this Letter we focus on the the photodissociation of homonuclear
molecules. Using co-propagating lasers, the momentum transfer associated with the
bound-free transition is negligible compared to the momentum spread of the BEC, i.e.
$\bfk\approx 0$. The `virtual cavity' modes then decouple into $\hat{c}_{1\bfq}$ and
$\hat{c}_{1-\bfq}$ pairs, resulting in a quadratic eigenvalue equation, with solutions
\begin{equation}
    \omega_\bfq=-\frac{i}{2}\gamma_\bfq
    \pm\sqrt{\frac{1}{4}\left[q^2-\delta\right]^2-|\chi|^2}.
\label{omegapm}
\end{equation}
The condition for positive exponential gain (${\mbox Im}\{\omega_\bfq\}>0$) is therefore
\begin{equation}
    \gamma^2_\bfq+\left[q^2-\delta\right]^2<4|\chi|^2,
\label{gaincond}
\end{equation}
leading to exponential growth of the mode occupation at the rate
\begin{equation}
    G_\bfq=2\sqrt{|\chi|^2-\frac{1}{4}\left[q^2-\delta\right]^2}-\gamma_\bfq.
\label{Gq}
\end{equation}

Equation (\ref{gaincond}) shows that bose-stimulation leads to exponential amplification
provided that the pumping is sufficiently strong to overcome the combined effects of losses
and de-phasing due to energy mismatch. Energy is conserved on a sphere in momentum space
centered at $\bfq=0$ with radius $\sqrt{\delta}$. The gain along this sphere, given by
$G_\bfq=2|\chi|-\gamma_\bfq$, attains a maximum when the directionally-dependent loss
$\gamma_\bfq$ is minimized. As a result, modes directed along the long-axis of the BEC will
grow the fastest.

The minimum loss rate value of $\beta\sqrt{\delta}$ along the condensate
axis, sets a power threshold of $2\chi\ge\beta\sqrt{\delta}$ for end-fire
amplification. In comparison, the stimulation threshold for side-fire modes,
$2\chi\ge\beta\alpha\sqrt{\delta}$ is larger by a factor of the aspect ratio $\alpha$. We
therefore identify three regimes
\begin{figure}
\begin{center}
\epsfig{file=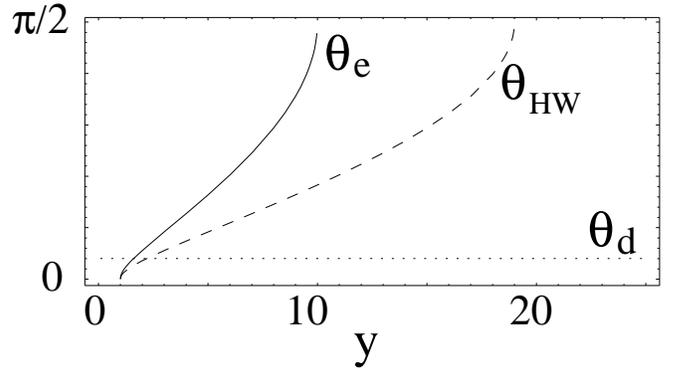,width=\columnwidth}
\end{center}
\caption{The exponential instability threshold angle $\theta_e$, the gain feature HWHM
angle $\theta_{HW}$, and the single-mode diffraction angle $\theta_d$ as functions of the
dimensionless control parameter $y$.} \label{f1}
\end{figure}
\noindent as the parameter $y=2\chi/(\beta\sqrt{\delta})$ is varied with respect to these
two thresholds :

{\bf Case I:} $y<1$. When the pump intensity is below both axial and radial thresholds,
there is simply isotropic spontaneous emission of atom pairs.

{\bf Case II:} $1<y< \alpha$. Between the thresholds, the exponential gain condition is
satisfied for modes with an emission angle $\theta_e$ from the z-axis. In general, a random
spot pattern will appear within $\theta_e$, with characteristic spot-size determined by the
angle subtended in k-space by a single resonant end-fire mode
$\theta_d\approx4\alpha/\sqrt{\delta}$.

Just above the longitudinal threshold ($1<y\ll\alpha$) we have $\theta_e\approx 1/\alpha$.
If $\theta_e$ is smaller than the spot-size $\theta_d$, only one end-fire mode in each
direction along the condensate axis will grow to macroscopic size. {\it This will be the
ideal situation for generating a number-squeezed pair of condensates}.

For $y$ extremely close to the lower threshold the exponential growth may be slow enough
that losses are not negligible during the build-up of the number-squeezed state. In this
case jets, i.e. out-coupled atom-laser beams, will be observed exiting the BEC volume along
the z-axis. This can lead to a decay of the number-squeezing, as the loss of atoms from the
end-fire modes is an uncorrelated random process. By slightly increasing the pump strength
slightly these can be made negligible without sacrificing the narrow emission angle.

As the pump intensity increases further into the intermediate regime, both the
width of the resonance shell and the solid angle of emission increase dramatically with $y$
due to power broadening . From Eq. (\ref{Gq}), we find that the angle with respect to the
BEC axis within which amplification can occur is
$\theta_e=\sin^{-1}\left[\frac{1}{\alpha}\sqrt{y^2-1}\right]$. Due to mode competition,
however, most emission will take place within the angle
$\theta_{HM}=\sin^{-1}\left[\frac{1}{2\alpha}\sqrt{(y+1)^2-4}\right]$, corresponding to the
half-maximum of the gain feature of Eq. (\ref{Gq}). In Figure 1 we plot $\theta_e$ and
$\theta_{HM}$ versus $y$, which shows both saturating at $\pi/2$ by $y=20$. The primary
result of this Letter, therefore, is that two-mode directionality occurs only in a narrow
operating regime $1<y<y_{cr}$, where $y_{cr}$ is the critical value where the emission
angle equals the diffraction angle $\theta_e=\theta_d$ In contrast, no such narrow
limitation exists in directional superradiant Rayleigh scattering \cite{inouyet,moore}.

{\bf Case III:} $y>\alpha$: When the pump intensity is above both thresholds, the emission
angle is comparable to the saturation angle $\pi/2$. While a slight gain differential may
still give some preference towards smaller angles, in the limit $y\gg\alpha$ the gain
differential becomes negligible, resulting again in isotropic emission.

In order to test the above predictions, based on the analysis of Eq. (\ref{Gq}), we solve
Eq. (\ref{pdqs}) numerically for the dynamics of the atomic field-operator $\hphi$. The
dominant quantum-field effect is amplification of spontaneously emitted atom pairs in the
vicinity of the instability. Therefore, Eq. (\ref{pdqs}) may be solved to a very good
approximation, by propagating its classical (mean-field) limit $\hphi_i\rightarrow\phi_i$,
with an initial atomic field seeded by random noise of order $1/\sqrt{N}$. We note that
this approach has been widely used to simulate parametric superfluorescence \cite{ewan} in
quantum optics. The increase in emission angle with $y$ is clearly seen in the
2-dimensional numerical simulations.

In Figure 2 we plot the atomic density in k-space for various values of the control
parameter $y$. The simulations use a Gaussian condensate wavefunction with aspect ratio
$\alpha=10$, a detuning $\delta=10^5$ and are integrated via a split-step FFT algorithm
until such time as $10^6$ atoms are emitted. The free Hamiltonian is taken to contain only
the standard kinetic energy term for simplicity. Adding a trap potential will have no
significant effect on the timescales we consider. In Figure 2a we show the case $y=2$, in
which a single pair of end-fire modes is observed. The case $y=10$ is shown in Fig 2b,
where we see that multiple modes are generated, all falling within the HWHM of Eq.
(\ref{Gq}), as predicted by our analysis. In figures 2c and 2d we plot the cases $y=20$ and
$y=100$, respectively, demonstrating the saturation of directionality as $y$ is increased
beyond the radial threshold. The contours in the figures correspond to our analytical
predictions as described in the caption. Quantitatively, the growth rate of the end-fire
modes, predicted by Eq. (\ref{Gq}) to be $G=2\chi-\beta\sqrt{\delta}$, was numerically
found to be $G=1.8\chi-5.4\sqrt{\delta}$, which corresponds to $\beta\approx 6$ for a
Gaussian wavepacket.

Interestingly, these simulations reveal the appearance of correlated pairs of modes which
closely resemble the BEC wavefunction shifted in momentum space, even though such
structures were not present in the initial noise which triggered the instability. This
lends additional validity to our choice of basis on which to analyze the
atomic field dynamics.

In conclusion, we analyze directional superradiant emission of atom
pairs from elongated BECs by constructing a simplified coupled-mode model which describes
both atom-photon and atom-atom pair generation. Our model indicates that while still
experimentally realizable, conditions for directional atom-atom pairs are significantly
more restrictive than for directional atom-photon pairs.

This work was supported by the National Science Foundation through a grant for the
Institute for Theoretical Atomic and Molecular Physics at Harvard University and
Smithsonian Astrophysical Observatory.

\begin{figure}
\begin{center}
\epsfig{file=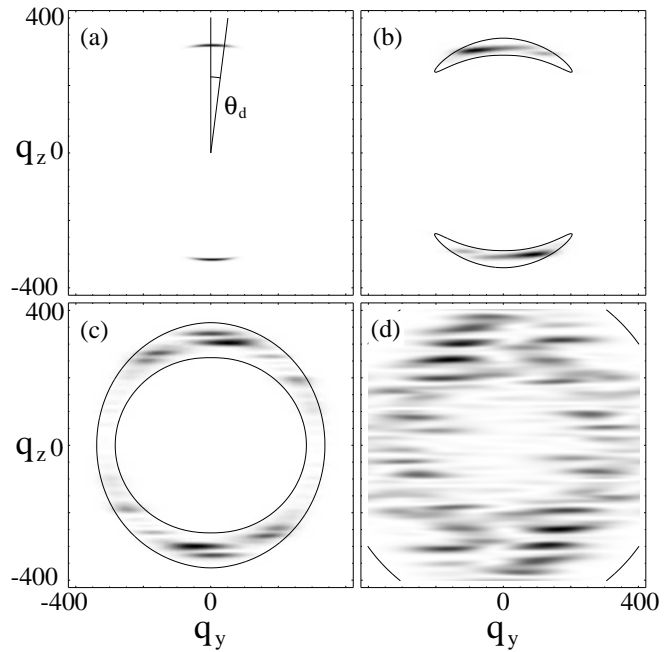,width=\columnwidth}
\end{center}
\caption{Numerical simulations of the atomic density in momentum space.
Atom number is $10^5$, $\alpha=10$ and $\delta=10^5$. Control parameter
values are (a) $y=2$, (b) $y=10$, (c) $y=20$, and (d) $y=100$. The single-mode
diffraction angle $\theta_d$ is indicated in (a), while in (b)-(d) the
contour corresponds to the half-maximum of the gain feature
of Eq. (\ref{Gq}).} \label{f2}
\end{figure}

\end{document}